\documentclass[aps,prc,twocolumn,epsf,groupedaddress]{revtex4}

\usepackage{graphicx}
\usepackage{dcolumn}
\usepackage{bm}

\begin{document}


\title{Measurements of electron-proton elastic cross sections for 
         0.4~$<$~$Q^2$~$<$~5.5~$({\rm GeV}/c)^2$}



\author{
M.E.~Christy,$^{7}$
A.~Ahmidouch,$^{15}$
C.~S.~Armstrong,$^{19}$
J.~Arrington,$^{2}$
R.~Asaturyan,$^{23}$
S.~Avery,$^{7}$
O.~K.~Baker,$^{7,19}$
D.~H.~Beck,$^{8}$
H.~P.~Blok,$^{21}$
C.~W.~Bochna,$^{8}$
W.~Boeglin,$^{4,19}$
P.~Bosted,$^{10}$
M.~Bouwhuis,$^{8}$
H.~Breuer,$^{9}$
D.~S.~Brown,$^{9}$
A.~Bruell,$^{11}$
R.~D.~Carlini,$^{19}$
N.~S.~Chant,$^{9}$
A.~Cochran,$^{7}$
L.~Cole,$^{7}$
S.~Danagoulian,$^{15}$
D.~B.~Day,$^{20}$
J.~Dunne,$^{13}$
D.~Dutta,$^{11}$
R.~Ent,$^{19}$
H.~C.~Fenker,$^{19}$
B.~Fox,$^{3}$
L.~Gan,$^{7}$
H.~Gao,$^{11}$
K.~Garrow,$^{19}$
D.~Gaskell,$^{2,17}$
A.~Gasparian,$^{7}$
D.~F.~Geesaman,$^{2}$
P.~L.~J.~Gu\`eye,$^{7}$
M.~Harvey,$^{7}$
R.~J.~Holt,$^{8}$
X.~Jiang,$^{17}$
C.~E.~Keppel,$^{7,19}$
E.~Kinney,$^{3}$
Y.~Liang,$^{1,7}$
W.~Lorenzon,$^{12}$
A.~Lung,$^{19}$
P.~Markowitz,$^{4,19}$
J.~W.~Martin,$^{11}$
K.~McIlhany,$^{11}$
D.~McKee,$^{14}$
D.~Meekins,$^{5}$
M.~A.~Miller,$^{8}$
R.~G.~Milner,$^{11}$
J.~H.~Mitchell,$^{19}$
H.~Mkrtchyan,$^{23}$
B.~A.~Mueller,$^{2}$
A.~Nathan,$^{8}$
G.~Niculescu,$^{16}$
I.~Niculescu,$^{6}$
T.~G.~O'Neill,$^{2}$
V.~Papavassiliou,$^{14,19}$
S.F.~Pate,$^{14,19}$
R.~B.~Piercey,$^{13}$
D.~Potterveld,$^{2}$
R.~D.~Ransome,$^{18}$
J.~Reinhold,$^{4,19}$
E.~Rollinde,$^{19,22}$
P.~Roos,$^{9}$
A.~J.~Sarty,$^{5}$
R.~Sawafta,$^{15}$
E.~C.~Schulte,$^{8}$
E.~Segbefia,$^{7}$
C.~Smith,$^{20}$
S.~Stepanyan,$^{23}$
S.~Strauch,$^{18}$
V.~Tadevosyan,$^{23}$
L.~Tang,$^{7,19}$
R.~Tieulent,$^{9,19}$
A.~Uzzle,$^{7}$
W.~F.~Vulcan,$^{19}$
S.~A.~Wood,$^{19}$
F.~Xiong,$^{11}$
L.~Yuan,$^{7}$
M.~Zeier,$^{20}$
B.~Zihlmann,$^{20}$
and V.~Ziskin$^{11}$.
}

email[]{christy@jlab.org}

\affiliation{
$^{1}${American University, Washington, D.C. 20016} \\
$^{2}${Argonne National Laboratory, Argonne, Illinois 60439} \\
$^{3}${University of Colorado, Boulder, Colorado 80309} \\
$^{4}${Florida International University, University Park, Florida 33199} \\
$^{5}${Florida State University, Tallahassee, Florida 32306} \\
$^{6}${The George Washington University, Washington, D.C. 20052} \\
$^{7}${Hampton University, Hampton, Virginia 23668} \\
$^{8}${University of Illinois, Champaign-Urbana, Illinois 61801} \\
$^{9}${University of Maryland, College Park, Maryland 20742} \\
$^{10}${University of Massachusetts, Amherst, MA 01003} \\
$^{11}${Massachusetts Institute of Technology, Cambridge, Massachusetts 02139} \\
$^{12}${University of Michigan, Ann Arbor, Michigan 48109} \\
$^{13}${Mississippi State University, Mississippi State, Mississippi 39762} \\
$^{14}${New Mexico State University, Las Cruces, New Mexico 88003} \\
$^{15}${North Carolina A \& T State University, Greensboro, North Carolina 27411} \\
$^{16}${Ohio University, Athens, Ohio 45071} \\
$^{17}${Oregon State University, Corvallis, Oregon 97331} \\
$^{18}${Rutgers University, New Brunswick, New Jersey 08855} \\
$^{19}${Thomas Jefferson National Accelerator Facility, Newport News, Virginia 23606} \\
$^{20}${University of Virginia, Charlottesville, Virginia 22901} \\
$^{21}${Vrije Universiteit, 1081 HV Amsterdam, The Netherlands} \\
$^{22}${College of William and Mary, Williamsburg, Virginia 23187} \\
$^{23}${Yerevan Physics Institute, 375036, Yerevan, Armenia} \\
}


\date{\today}

\begin{abstract}
We report on precision measurements of the elastic cross section for electron-proton 
scattering performed in Hall C at Jefferson Lab.  The measurements were made at 28 distinct 
kinematic settings covering a range in momentum transfer of $0.4< Q^2$ $<$ 5.5 $({\rm GeV}/c)^2$.
These measurements represent a significant contribution to the world's
cross section data set in the $Q^2$ range where a large discrepancy currently exists
between the ratio of electric to magnetic proton form factors extracted from previous
cross section measurements and that recently measured via polarization transfer in Hall A
at Jefferson Lab.  This data set shows good agreement with previous cross section 
measurements, indicating that if a here-to-fore unknown systematic error does exist in the 
cross section measurements then it is intrinsic to all such measurements.
\end{abstract}


\maketitle


\section{Introduction}

Recently, there has been much renewed interest in the proton electromagnetic form factors in the 
region of four-momentum transfer, $Q^2$ $>$ 1 $({\rm GeV}/c)^2$.  This is due primarily to 
recent measurements from Hall A at Jefferson Lab~\cite{jones,gayou1} on the ratio of the Sachs 
electric to magnetic form factors via the polarization transfer technique~\cite{pol1,pol2}.  
These data are in stark disagreement with previous extractions of these form 
factors~\cite{litt,price,walker,andiv} from cross section measurements utilizing the 
Rosenbluth separation technique~\cite{rosen}.  

There have been recent efforts~\cite{brash,arrington,arrington2} to 
extract the individual form factors by combining the cross section and polarization transfer 
results.  However, it is clear that the data sets from these two 
techniques are systematically inconsistent~\cite{arrington} and, as such, the method that 
one chooses for combining the data sets is not well defined.  It is critical, then, that the 
source of the discrepancy be identified, if there is to be any chance of pinning down the $Q^2$ 
dependence of the individual form factors~\cite{arrington2}.

In this paper we will present results from 28 new precision measurements of the {\it ep} elastic 
cross section in the range 0.4~$<$~$Q^2$~$<$~5.5~$({\rm GeV}/c)^2$ performed in Hall C at Jefferson 
Lab.  Although the kinematics are such that only limited Rosenbluth separations of the form 
factors can be performed, this data represents a significant contribution to the world's cross 
section data set, and as such, can help provide new constraints on global fits from which the form 
factors can be extracted. 

 The high precision and large kinematic coverage of the new Jefferson Lab data can help provide 
crucial information as to whether there exists an experimental systematic error in the world's 
cross section data set, which is dominated by the data from SLAC in the $Q^2$ range where the 
discrepancy with the polarization transfer data exists.

\section{{ep} Elastic Scattering}

\begin{figure}
\includegraphics[width=6cm,height=4cm]{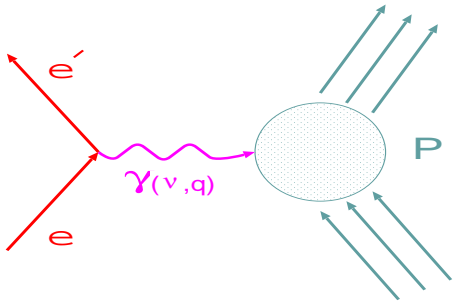}
\caption{\label{opea} (Color online) Single photon exchange diagram for electron proton elastic scattering.}
\end{figure}   

The elastic scattering of an electron from a proton target can be represented in the
first order Born approximation by the exchange of a single virtual photon between the 
leptonic and hadronic electromagnetic currents.  This exchange is represented by the 
diagram in Fig.~\ref{opea} and is often referred to as the one photon exchange 
approximation (OPEA), with 4-momentum transfer
\begin{equation}
 q_{\mu} = k'_{\mu} - k_{\mu}, 
\label{qdef}
\end{equation}
where $k_{\mu}$ ($k'_{\mu}$) is the four-momentum of the electron before (after) scattering.
For space-like photons ($q^2 = q_{\mu}q^{\mu}$ $<$ 0) it is customary to define the absolute 
value of the square of the four-momentum transfer 
\begin{equation}
Q^2 \equiv -q^2 \approx 4EE'\sin^2(\theta/2).
\label{q2def}
\end{equation}

If the proton were point-like, then the cross section could be calculated within the 
framework of quantum electrodynamics (QED) to give
\begin{equation}
\sigma_{pl} \equiv \frac{d\sigma_{pl}}{d\Omega} = \frac{E'}{E} \frac{\alpha^2 \cos^2(\theta/2)}
{4E^2 \sin^4(\theta/2)}.
\label{eq:cs1}
\end{equation}
However, the spatial extent of the electromagnetic charge and current densities of the proton 
lead to the introduction of form factors, which modify the proton vertex and parameterize the 
protons internal structure.  
It is common to see the cross section expressed in terms of the Sachs electric and magnetic 
form factors, ${G_E}_p$ and ${G_M}_p$.  These form factors are defined in such a way that 
only terms quadratic in them appear in the Rosenbluth expression for the cross section,

\begin{equation}
\frac{d\sigma}{d\Omega} = \sigma_{pl} \left[ \frac{{G_E^2}_p(Q^2) + \tau {G^2_M}_p(Q^2)}{1 + \tau} 
      + 2\tau {G^2_M}_p(Q^2)\tan^2(\theta/2)\right],
\label{eq:cs3}
\end{equation}
where $\tau \equiv Q^2/4M_p^2$ and $M_p$ is the proton mass.  In the non-relativistic limit, 
${G_E}_p$ is given by the Fourier 
transform of the spatial charge distribution, while ${G_M}_p$ is given by the Fourier transform 
of the spatial magnetization distribution.  At zero momentum transfer, the proton is resolved as a point 
particle of total charge equal to one and total magnetic moment given by
 $\mu_p = 1  + \kappa_p$, 
where $\kappa_p = 1.7928$ is the proton anomalous magnetic moment. 
This leads to the normalizations 
\begin{equation}
{G_E}_p(0) = 1 \text{ \rm and } {G_M}_p(0) = \mu_p. 
\end{equation}

\section{Extraction of Form Factors From Cross Section Measurements}

\begin{figure}
\includegraphics[width=8cm,height=8cm]{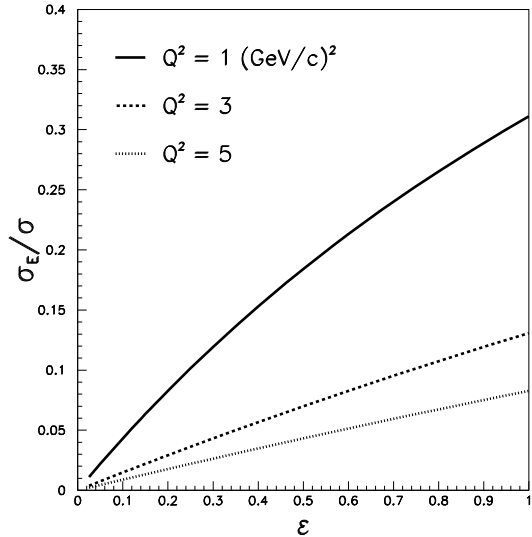}
\caption{\label{ffsens} Fractional contribution of $G_E$ to the cross section assuming 
${G_M}/ \mu_p = G_E$ (form factor scaling).}
\end{figure}

The Rosenbluth expression, Equation~\ref{eq:cs3}, can be recast in terms of the relative 
longitudinal polarization of the virtual photon, 
$\varepsilon = \left[1 + 2(1 + \tau)\tan^2(\theta/2)  \right]^{-1}$, as 
\begin{equation}
\frac{d\sigma}{d\Omega} = \frac{\sigma_{pl}}{\varepsilon (1 + \tau)} \left 
[ {\varepsilon {G_E^2}_p(Q^2) + \tau {G^2_M}_p(Q^2)} \right ],
\label{eq:cs4}
\end{equation}
with the reduced cross section defined by 
\begin{equation}
\sigma_{r} \equiv \frac{d\sigma}{d\Omega} \cdot \frac {\varepsilon (1 + \tau)}{\sigma_{pl} } 
= {\varepsilon {G_E^2}_p(Q^2) + \tau {G^2_M}_p(Q^2)}.
\label{eq:cs5}
\end{equation}
At fixed $Q^2$, the individual form factors, $G_E$ and $G_M$, can be extracted from a linear fit 
in $\varepsilon$ to the measured reduced cross section.  Such a fit is generally referred to as a 
Rosenbluth fit and yields $\tau G^2_M$ as the intercept and $G^2_E$ as the slope.  Due to the 
$\tau$ weighting of $G^2_M$, the cross section becomes less sensitive to $G_E$ at large $Q^2$.  
Hence, the accuracy with which $G_E$ can be extracted decreases (inversely) with $Q^2$ and Rosenbluth 
separations eventually fail to provide information on the value of $G_E$.  
This failure was part of the impetus for the development of the polarization transfer technique.  
The fractional contribution of $G_E$ to the cross section assuming ${G_M}/ \mu_p = G_E$ 
(form factor scaling) is shown as a function of $\varepsilon$ in Fig.~\ref{ffsens} for $Q^2$ values 
of 1, 3, and 5 $({\rm GeV}/c)^2$.  At $Q^2$ = 3 $({\rm GeV}/c)^2$, $G_E$ contributes only 12$\%$ to 
the cross section at $\varepsilon$ = 1, with this contribution decreasing approximately linearly 
as $\varepsilon \rightarrow$ 0 in this $Q^2$ range. 

\section{Experiment}

The {\it ep} elastic scattering data presented here were obtained as part of experiment 
E94-110~\cite{e94110}, which was intended to separate the longitudinal and transverse 
unpolarized proton structure functions in the nucleon resonance region via Rosenbluth separations.  
The experiment utilized the high luminosity electron beam provided by the CEBAF accelerator and was 
performed in Jefferson Lab Hall C during summer and fall of 1999.  Scattered electrons were detected 
in the High Momentum Spectrometer (HMS).  Additionally, the Short Orbit Spectrometer (SOS) was used 
to detect positrons, which were used to determine possible electron backgrounds originating from 
charge-symmetric processes such as $\pi^0$ production and subsequent decay in the target.  For the 
kinematic of the elastic scattering measurements, these backgrounds were found to be less than $0.1\%$.

\subsection{Hall C Beamline}

\begin{figure}
\includegraphics[width=8cm]{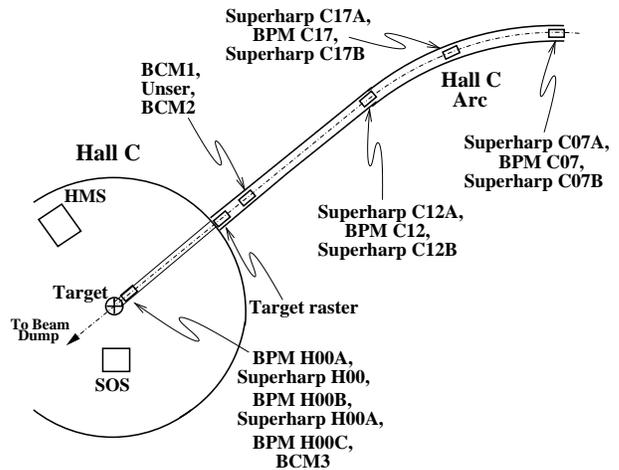}
\caption{\label{hallcbeamline} Schematic of the Hall C beamline.}
\end{figure}

The Hall C beamline from the beam switch yard to the beam dump in the experimental area is shown in 
Fig.~\ref{hallcbeamline}.  The beam from the accelerator south linac enters the Hall C Arc and 
passes through a series of dipole and quadrupole magnets which steer it into the Hall.  The beam 
position and profile can be measured at several stages in the Arc with the use of superharps.  The
superharps consist of a set of fine wires (two horizontal and one vertical) which are moved back 
and forth through the beam to determine the centroid position to about 10 $\mu m$.  However, these 
measurements are invasive and can not be performed during data taking.  Continuous monitoring of the 
beam position in the Arc is done with the aid of three beam position monitors (BPMs), which are 
nondestructive to the beam and are calibrated with superharp scans.  

The absolute beam position provided by scans 
of each of the three superharps allows the trajectory of the beam through the magnets to be determined.  
This, combined with knowledge of the field integrals of the Arc magnets, then allows the absolute beam 
energy to be determined to better than $0.1\%$.  Absolute beam energy measurements which require superharp 
scans were performed about twice per beam energy setting.
 
Accelerator cavity RF instabilities have been observed to cause variations in the beam energy of 
about 0.05$\%$.  These variations of the beam energy can be measured using the relative positions 
provided by the Arc BPMs.  These BPMs were read into the data stream every second and used to monitor 
the beam energy drift.  In principle, the effect of a drift can be corrected for if a large enough 
sample of events is considered.  However, the effect of beam energy drift on runs used in the current 
analysis was studied and found to be less than a $0.02\%$ effect on the beam energy.

The beam position monitoring system in the Hall consists of three BPMs and two superharps for calibrations.  
Deviations in the angles of the beam on target translate into corresponding offsets in the reconstructed angles, 
whereas deviations in the vertical (spectrometer dispersive direction) position of the beam will manifest themselves a
s apparent momentum and out-of-plane angle offsets in the spectrometers.   The effect of a beam position offset 
can be calculated from the optical matrix elements for the spectrometer.  For a 1 mm vertical offset of 
the beam on target, the shifts in the reconstructed momentum and out-of-plane angle in the HMS are 
about $0.08\%$ and 1~mrad, respectively.     

The centroid of the beam spot, determined by the beam steering into the Hall, is constantly monitored by 
both fast-feedback electronics and visual displays of the BPM readouts and is adjusted to prevent large 
drifts of the on-target position during data taking.  A study of the run-to-run beam steering stability was 
made during the running of this experiment.  In this study, the run-to-run variations in the vertical position 
on target were measured to be less than 0.2 mm, resulting in a corresponding point-to-point uncertainty in the 
reconstructed momentum of $0.016\%$. The run-to-run variations in the angles on target were found to be 
less than 0.04~mrad.

In order to minimize localized target boiling effects in the liquid hydrogen, the small intrinsic beam spot 
size of about 300 $\mu \rm m^{1}$\footnotetext[1]{This is about 3$\times$ the normal intrinsic spot size.} was 
increased by a set of fast rastering magnets before entering the Hall.  
The fast raster produced a rectangular pattern with a full width of about 4 mm in the horizontal and 2 mm in 
the vertical.  Corrections due to the vertical rastering were calculated and corrected event-by-event.

The beam current monitoring system in the Hall consists of two (three for E94-110) resonant microwave cavity 
beam current monitors (BCMs).  The BCMs provide continuous measurement of the current and are calibrated to 
about 0.2 $\mu$A by use of an Unser monitor in the Hall.  Dedicated calibration runs were performed about once 
every three days during this experiment to minimize the effects of drifts in the BCM gains.  The current was 
carefully monitored during data taking and was required to be $60 \pm 2 \mu$A.  The normalization uncertainty 
due mostly to the Unser was estimated to be 0.4\%.   The run-to-run uncertainty in the beam current of 0.2\% 
was estimated by combining in quadrature the fit residuals from the calibration runs and the typical observed 
drift between calibrations.  Detailed information on the current monitoring systems in Hall C can be found in 
Reference~\cite{armstrong}.

\subsection{Target}

\begin{figure}
\includegraphics[width=8cm]{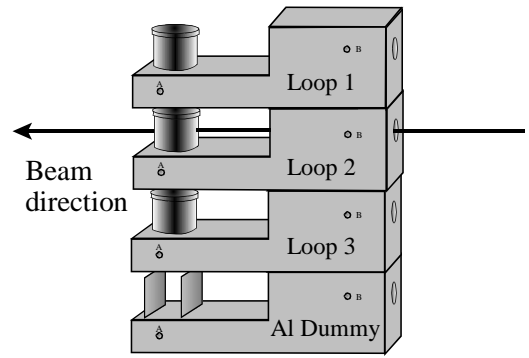}
\caption{\label{target} Representation of the cryogenic and dummy target assembly.}
\end{figure}

A representation of the cryogenic target assembly is displayed in Fig.~\ref{target}, and shows the three 
``tuna can'' shaped cryogen cells, as well as the dummy target.  Each can was machined out of aluminum to 
provide a very uniform cylindrical shape which 'bulges' a negligible amount when the cell is 
pressurized to about 25 psia~\cite{dunne}.  The hydrogen cell was measured to have an inside 
diameter of 40.113 mm when warm and 39.932 mm when cold, and a cylindrical wall thickness of 0.125 mm.  
Due to the circular shape, the average target length seen by the beam depended upon both the central position of 
the beam spot and the size and form of the raster pattern.  The normalization uncertainty in the hydrogen 
target length was estimated to be 0.3\% and the run-to-run uncertainty was estimated to be 0.1\%.  
The dummy target was made from two 0.975 mm thick, rectangular sheets of aluminum separated by 40 mm.  Additional 
details on the target assembly can be found in Reference~\cite{dunne}.

Localized target density fluctuations, induced by an intense incident beam, can modify significantly the
average density of a cryogenic target.  Uncertainties in target density enter directly
as uncertainties in the total cross section, and can be current-dependent on a point-to-point basis.  The
current-dependence can be measured by comparing the yields at fixed kinematics with varying beam currents.
The deadtime-corrected yields should be proportional to the luminosity (and, therefore, to the target density).

\begin{figure}
\includegraphics[width=8cm,height=7cm]{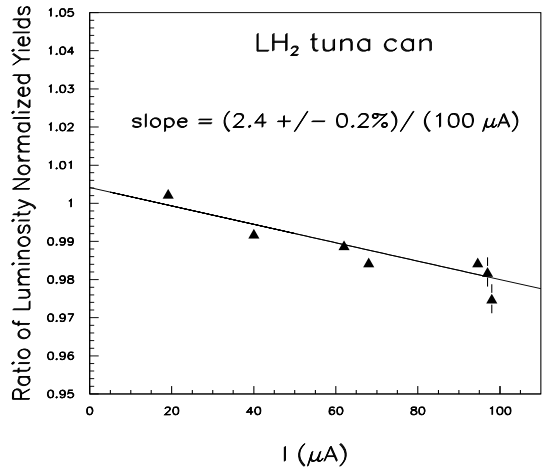}
\caption{\label{boiling} Relative hydrogen target yield versus beam current.}
\end{figure}

The result of such a `luminosity scan' for E94-110 is shown in Fig.~\ref{boiling}, where the luminosity
relative to the lowest current has been plotted on the vertical axis. The error bars on the data are 
statistical only and do not reflect fluctuations in the beam current.  The correction factor applied to the 
measured target density (at zero current) to account for the reduction resulting from 
localized target boiling is given by the product of the fitted slope and the current at which the data was 
taken.  For the present data the current was typically kept at 60 $\pm$ 2 $\mu$A, resulting in a density 
correction of 
\begin{equation}
(2.4 \pm 0.2)\% \cdot 60\mu {\rm A}/ 100\mu {\rm A} = 1.44 \pm 0.12\%.
\end{equation}  
The uncertainty in the current did not contribute appreciably to the uncertainty on this correction.  The 
total estimated run-to-run uncertainty in the target density is 0.1\%. 

\subsection{HMS Spectrometer}

\begin{figure} 
\includegraphics[width=8cm,height=6cm]{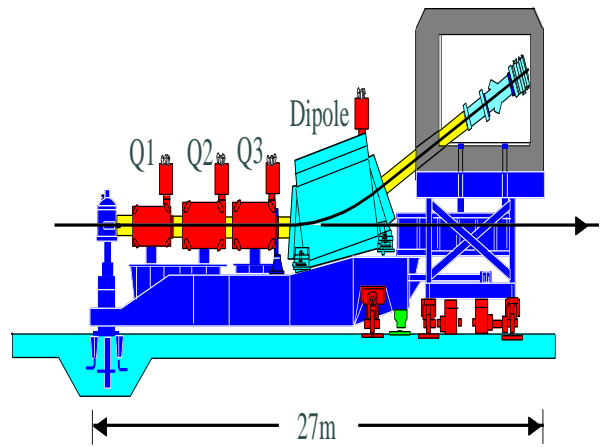}
\caption{\label{hms_full} (Color online) Schematic drawing of the HMS spectrometer.}
\end{figure}

The HMS is a magnetic spectrometer consisting of a $25^{\circ}$ vertical bend dipole 
magnet (D) for momentum dispersion and three quadrupole magnets (Q1, Q2, Q3) for focusing.  
All magnets are superconducting and were operated in a mode to provide a point-to-point 
optical tune.  A schematic side view of the HMS is shown in Fig.~\ref{hms_full}, and includes 
representations of the pivot (with target chamber), magnets, and the shielded hut containing 
the detector stack.

\begin{figure}
\includegraphics[width=8.5cm,height=4cm]{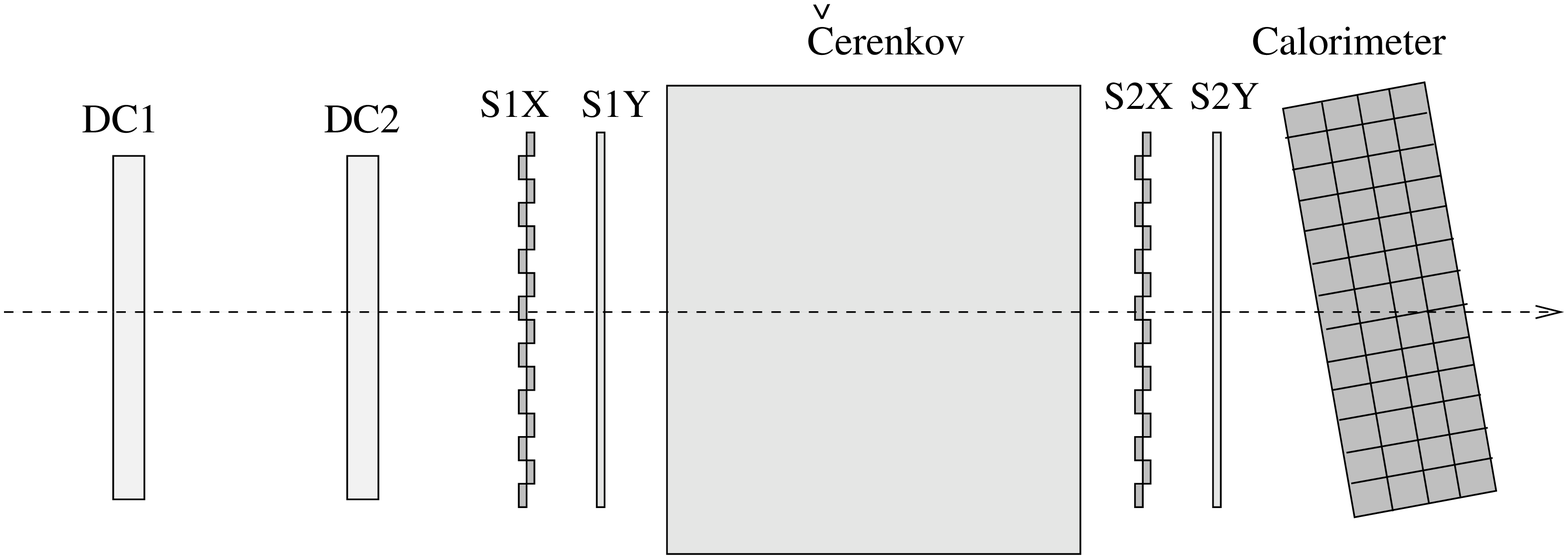}
\caption{\label{hms_detector} Schematic drawing of the HMS detector stack.  The 
component detectors are described in the text.}
\end{figure}

The detector stack is shown in Fig.~\ref{hms_detector} and consists of two vertical drift 
chambers~\cite{baker} (DC1 and DC2) for track reconstruction, scintillator arrays (S1X(Y) and S2X(Y)) for 
triggering, and a threshold gas \v Cerenkov and electromagnetic calorimeter, which were both used in the 
present experiment for particle identification (PID) and pion rejection.  

The acceptance limits of the HMS in-plane ($Y'$) and out-of-plane ($X^{\prime}$) scattering angles are 
defined by an octagonal collimator positioned between the target and the first quadrupole magnet.  The edges 
of this collimator define a maximum angular acceptance of $-28~<~Y^{\prime}~<~28$~mrad and 
$-75~<~X^{\prime}~<~75$~mrad, and a total solid angle of about 6.8~msr.  Additional details on the HMS can 
be found elsewhere~\cite{dutta}.

\subsection{Data Acquisition}

Data acquisition was performed using the CEBAF On-line Data Acquisition (CODA) software~\cite{abbot1} 
running on a SUN Ultra-2 workstation.  The detector information for each event was collected from the front-end 
electronics by VME/CAMAC computers (collectively referred to as Read-Out Controllers or ROCs).  Event fragments 
from the ROCs were then transfered via TCP/IP to CODA, which formed events and wrote them to disk.

\section{Data Analysis}

CODA events from individual run files where decoded by the Hall C Replay software, which reconstructed the 
trajectories of individual particles from hit information in the drift chambers.  Tracks were then transported 
back to the target via an optical transport model of the HMS, which allowed the determination of the particle 
kinematics.   For each run an HBOOK~\cite{hbook} ntuple was then created which contained the reconstructed event 
kinematics and calibrated PID detector information.  The final analysis of the ntuples into experimental yields 
is described in the sections which follow.

\subsection{Kinematic Calibrations}

One of the larger $\varepsilon$-dependent uncertainties that directly affects Rosenbluth separations is 
that due to the uncertainties in the kinematics at which the cross sections are measured.  It is 
convenient to absorb this uncertainty directly into the cross sections by calculating the expected 
difference in the measured cross section when the kinematics are changed from the nominal values within 
their uncertainties.  In order to minimize this uncertainty, it was critical that the kinematic 
quantities, $E$, $E^{\prime}$, and $\theta$ be determined to the best possible precision.  This was aided by 
the kinematic constraint of elastic scattering, that the reconstructed mass of the unmeasured hadronic 
state be equal to the proton mass.
        
For each kinematic setting, the difference of the reconstructed invariant mass, W, from the proton
mass ($\Delta W$ = $W$ - $M_p$) was calculated after correcting for the effects of energy loss due to 
both ionization and bremsstrahlung emission.  This provided a large set of kinematics for which the 
dependence of $\Delta W$ on possible energy and angle offsets could be studied.  Finally a minimization 
of $\Delta W$ was performed to determine the best set of kinematic offsets under the following assumptions: 
1) the offset of the nominal HMS central momentum from the true value was a constant fractional amount, 
and 2) the offset of the nominal HMS central angle from the true value was a constant.  The nominal 
HMS momentum used in this study was that determined in ~\cite{dutta}, while the nominal HMS central angle 
was determined from a comparison of marks scribed on the floor of the Hall to a marker on the back of the 
spectrometer, which indicated the optical axis.          

\begin{figure}
\includegraphics[height=10cm,width=8cm]{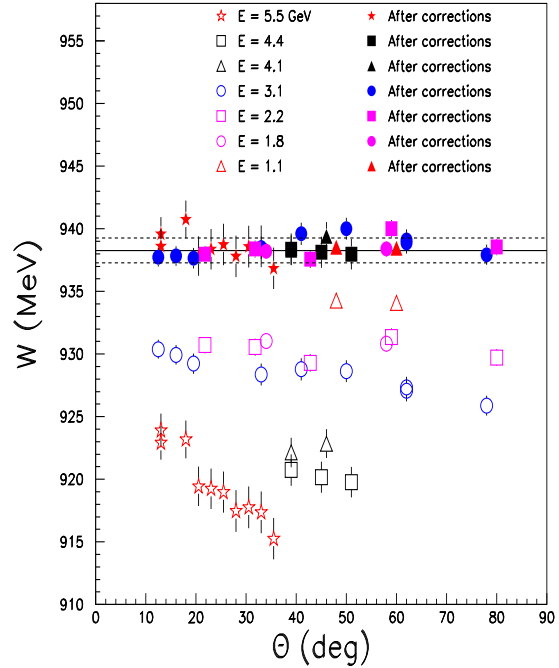}
\caption{\label{offsets} (Color online) Reconstructed $W$ vs. HMS central angle for elastic 
scattering kinematics.  Open symbols represent data before kinematic corrections were applied, 
while the solid symbols represent the data after applying the calibration corrections.}
\end{figure}

The reconstructed $W$ values for these data are plotted versus scattering energy in 
Fig.~\ref{offsets}, for eight different beam energies and thirty one unique kinematic 
settings.  It was found that the entire data set could be well described by assuming 
that the true HMS central angle was smaller than the nominal value by 0.6 mrad, and 
that the true HMS central energy was smaller than the nominal value by 0.39$\%$.  

The true 
beam energy was also found to be smaller than the Arc measurements by an amount that 
varied with the energy.  This latter result was subsequently confirmed~\cite{mack} by a 
remapping and analysis of the field for one of the Arc magnets.  The reconstructed values 
for $W$ are shown, both before (open symbols) and after (solid symbols) correcting for the 
kinematic offsets found from these studies.  The corrected values are all seen to be 
within 1-2 MeV of the proton mass.  This procedure was used for the E94-110 data and 
yielded a estimated uncertainty in the corrected beam energy of 0.056$\%$, about half that 
typically quoted from Hall C Arc measurements.  We estimate that this uncertainty is the 
quadrature sum of equal normalization and run-to-run uncertainties.  We note that the beam 
energies determined from the Arc measurements utilizing the updated field maps agree with 
the current results to about 0.05\%.

\subsection{Binning the Data}

The data were binned on a 2-dimensional grid in the reconstructed variables 
$E^{\prime}$ and $\theta$.  This was because, at fixed beam energy, the inclusive cross 
section only depends on the scattered electron energy and angle.  In practice, the binning 
in $E^{\prime}$ was converted to a binning in $\delta P/P$, which is the more natural variable 
for the application of the acceptance corrections.  The ranges were chosen such that the entire 
angular acceptance was included and the $\delta P/P$ acceptance was well determined from the 
model of the HMS.  For $\delta P/P$, the binning chosen was 16 bins over a range of $\pm 8\%$, while 
for $\theta$ the binning chosen was 20 bins over a range of $\pm 35$ mrad 
($\Delta \theta$~=~3.5~mrad).  We note that the physical solid angle coverage ($\Delta \Omega$) 
can be different for each $\Delta \theta$.

\subsection{Analysis Procedure}
For a beam of electrons of energy $E$ incident on a fixed proton target, the number of electrons 
scattered at an angle $\theta$ in a solid angle $\Delta\Omega$ is related to the differential 
cross section, ${d\sigma(\theta)/d\Omega}$ by
\begin{equation}
N(\theta) = \mathcal{L} \cdot \frac{d\sigma(\theta)}{d\Omega}\Delta\Omega,
\end{equation} where $\mathcal{L}$ is the integrated luminosity.   This is not the OPEA cross 
section of Equation~\ref{eq:cs4}, but contains contributions from higher order QED effects.  These 
include virtual particle loops, multi-photon exchange, as well as the emission of bremsstrahlung photons, 
both before and after the scattering.  

The emission of unmeasured bremsstrahlung photons by either the electron beam or the outgoing 
detected electron results in energies at the scattering vertex which are either smaller (the incoming case) 
or larger (the outgoing case) from those used in the reconstruction of the kinematics.  
This results in a large radiative ``tail'' in both the reconstructed $E^{\prime}$ and the invariant hadron 
energy distributions for elastic events.  To compare to the OPEA cross section, requires that this radiative 
tail be integrated to some cut-off in $E^{\prime}$, with a correction factor, which included the remaining higher 
order effects, depending on this cut-off.  This ``radiative'' correction is applied as a multiplicative factor 
(denoted RC) and is discussed in more 
detail in section \ref{sect:rc}.  Because the radiative tail extends beyond the threshold for single pion 
production at $W^2 \approx 1.16$ $\rm GeV^2$, the integration was cutoff at $W^2_{max} < 1.16$ $\rm GeV^2$ 
to avoid including events from inelastic processes.  The corresponding correction factor, RC($W^2_{max}$), is 
therefore cutoff dependent.

In addition, the measured number of counts must also be corrected for detector efficiencies, $eff$, and 
the effective solid angle acceptance, $\Delta\Omega_{eff}(\theta,E^{\prime})$, after subtraction of counts from 
background processes, BG$(\theta,E^{\prime})$.  In this experiment the measured cross section was determined 
for each bin on a 2-dimensional grid of the electron scattering energy and angle, $E^{\prime}$ and $\theta$, 
across the entire phase space for which the spectrometer has a non-zero acceptance.
\noindent
The extracted cross section was then determined from the relation,
\begin{equation}
\frac{d\sigma_{1\gamma}(\theta)}{d\Omega} = \frac{RC(W^2_{max})}{\mathcal{L}}{\int}^{W^2_{max}} 
{\it dE^{\prime}} 
\frac{[N(E^{\prime},\theta) - BG(E^{\prime},\theta)]}{eff \cdot \Delta\Omega_{eff}(E^{\prime},\theta)}.
\end{equation}
The individual ingredients will be discussed in detail in the following sections.

\subsection{Backgrounds}

There are three physical processes that are possible sources of background counts to the elastically 
scattered electron yields.  These are:  electrons scattered from the target aluminum walls, 
negatively charged pions that are not separated from electrons by the PID cuts, and electrons originating 
from other processes, which are dominated by charge symmetric processes which produce equal numbers of 
positrons.  Each of these potential backgrounds will be examined in the discussion that follows.

\subsubsection{Target Cell Backgrounds}

The quasielastic scattering from nucleons in aluminum nuclei can produce electrons at the same kinematics 
as those from elastic {\it ep} scattering.  The scattering of the beam from front and back of the target 
cell wall produces backgrounds of this type which are difficult to isolate.  Therefore, the corresponding 
background is determined by measuring the yield of events from a ``dummy'' target, which is a mockup of the 
target ends.  In order to minimize the data acquisition time, the total thickness of this dummy target was 
about 8 times the total cell wall thickness seen by the beam.  After measuring the dummy yield, the total 
background from scattering from the target walls, ${\rm BG}_{w}(E^{\prime},\theta)$, was then determined from 
\begin{equation}
{\rm BG}_{w}(E^{\prime},\theta) = \frac{t_{w} Q_{w} }{t_{d} Q_{d}}N_{d}(E^{\prime},\theta)\cdot 
C_{\rm br}(E^{\prime},\theta),
\end{equation}
where $Q_{w(d)}$ is the total charge incident on the walls (dummy), $t_{w(d)}$ is the total thickness of 
the walls (dummy), and $N_d(E^{\prime},\theta)$ is the number of events collected for the dummy run after applying 
efficiency and deadtime corrections. 

The factor, $C_{\rm br}(E^{\prime},\theta)$, corrects for the difference in external bremsstrahlung emission 
due to the greater thickness of the dummy target.  More precisely, this accounts for the fact that the 
distribution of events for thicker targets are more strongly shifted toward lower scattering energies 
(higher $W$) than those for thinner targets.  The size of this correction was studied and was found to be 
less than a few tenths of a percent at all kinematics measured, and typically less than $0.1\%$.  Since this 
was the typical size of the uncertainty in this correction, we have taken $C_{\rm br} = 1$, and absorbed an 
additional $0.1\%(0.1\%)$ into the point-to-point (normalized) uncertainty in the background subtraction. 

The largest contribution to the uncertainties in the aluminum background subtraction comes from the 
uncertainties in the thickness of the cell wall of about $ 1.5\%$~\cite{dunne}.  However, the typical size of 
this background was on the order of $8\%$ of the total yield, which leads to an uncertainty on the 
subtracted yield of only $0.12\%$.  This uncertainty is approximately independent of the kinematics and run 
conditions.

\subsubsection{Pion Backgrounds}

\begin{figure}
\includegraphics[height=9cm,width=8cm]{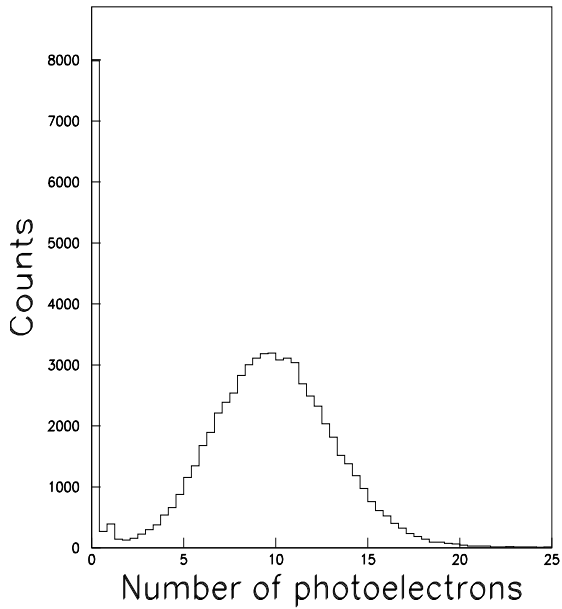}%
\caption{\label{cer_pub}Distribution of the number of photoelectrons collected 
in the \v Cerenkov detector for elastic kinematics of $E$ = 2.2 GeV and $p$ = 1 GeV/$c$.} 
\end{figure}

\begin{figure}
\includegraphics[height=9cm,width=8cm]{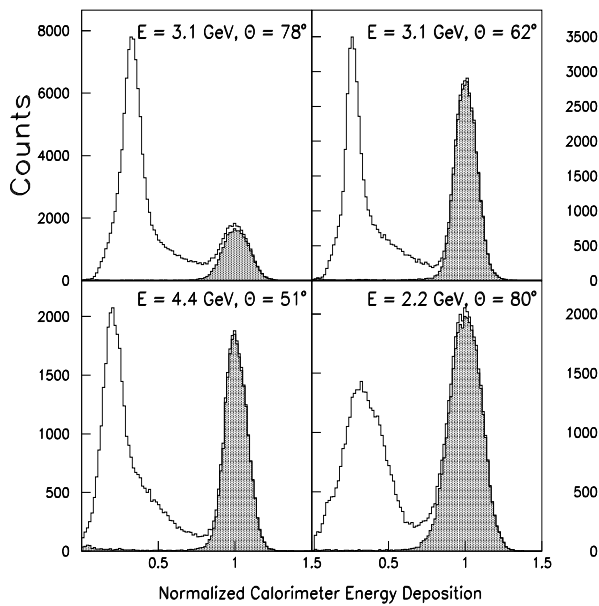}
\caption{\label{cal_pub}Distribution of the fractional energy deposition in the calorimeter for 
events both before (open area) and after(shaded area) applying a requirement of $> 2$ photoelectrons 
in the \v Cerenkov detector.} 
\end{figure}

The rejection of negatively charged pions was accomplished by placing requirements on both the number of 
\v Cerenkov photoelectrons collected and the energy deposition of the particle in the calorimeter.  
The count distribution of photoelectrons collected in the HMS \v Cerenkov at an HMS momentum of 1 GeV is shown 
in Fig.~\ref{cer_pub}.  For electrons, this is a Poisson distribution, with a mean of approximately 10 
photoelectrons.  For pions, the number of photoelectrons produced should be zero.  However, pions 
can produce $\delta$-rays (electron knockout) in the materials immediately preceding the \v Cerenkov detector 
and some of these ``knock-on'' electrons can produce \v Cerenkov radiation, with the probability of 
$\delta$-ray production increasing with energy.  With a requirement of more than 2 photoelectrons, this 
decreases the pion rejection factor from the maximum value of about $1000:1$ found at low energies.  However, 
this doesn't cause any significant pion contamination above this cut since the worst $\pi^-/e^-$ ratios are at 
low scattering energy where the rejection factor is the largest.

The fractional energy deposition in the calorimeter, both before (unshaded region) and after (shaded region) 
applying the \v Cerenkov requirement of $>2$ photoelectrons to select electrons, is shown in Fig.~\ref{cal_pub} 
for the four kinematics that exhibited the worst $\pi^-/e^-$ ratio.  The fractional energy deposition of the 
particles is calculated by dividing the energy collected in a fiducial region about the track in the 
calorimeter by the momentum determined from the track reconstruction.  Even for these worst cases, it is 
evident that the \v Cerenkov requirement alone does a good job of removing pions.  To further insure a clean 
electron sample, a requirement that the fractional energy deposited in the calorimeter be greater than $0.7$ 
was also applied.  The pion background after applying both the \v Cerenkov and calorimeter requirements 
is estimated to be less than $0.1\%$. 

The \v Cerenkov efficiency, using a 2 photoelectron cut, was found to be $99.6\%$, independent of 
the energy.  This is because the shape of the electron distribution does not depend on the particle's energy, 
resulting in the same fraction of electrons in the tail being removed.  This is not true for the calorimeter, since 
a fixed energy resolution results in an increase in the width of the electron fractional energy distribution at 
lower energies.  The 
calorimeter cut efficiency decreases from a maximum of about $99.5\%$ at energies above 3 GeV to about $98.5\%$ at 
an energy of 0.6 GeV.  The run-to-run uncertainties on the efficiencies were estimated from Gaussian fits 
of the distribution of efficiencies determined for each run from the entire E94-110 elastic data set, and were 
found to be $0.1\%$ for the \v Cerenkov detector and $0.1\%$ for the calorimeter.

\subsection{Acceptance Corrections}

\begin{figure}
\includegraphics[height=9cm,width=8cm]{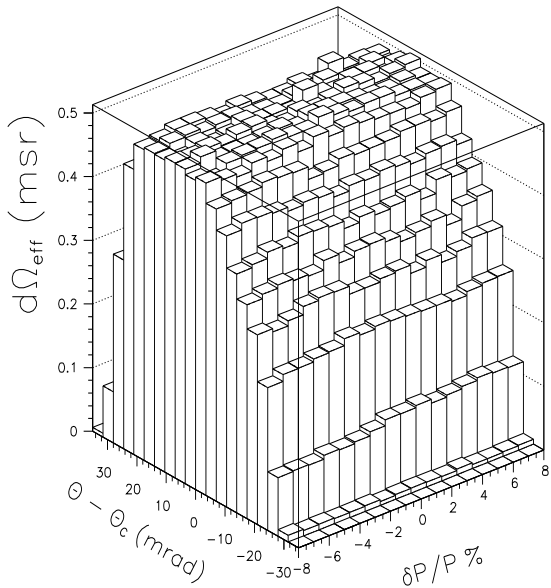}%
\caption{\label{accpt}HMS effective solid angle plotted in $\theta - \delta P/P$ space, using 
the binning described in the text.}
\end{figure}

Whether a scattered electron reaches the detector stack or is stopped by hitting the edge of the collimator 
or one of the various apertures in the HMS magnet system and beam pipe is dependent upon several factors, including:  
1) the electron momentum, 2) the in-plane and out-of-plane scattering angles, and 3) the vertex position.  
However, the physics depends only upon the momentum and full scattering angle, 
$\theta = \cos^{-1}\left[\cos(X^{\prime})\cos(\theta_c-Y^{\prime})\right]$, so that for a fixed central 
spectrometer angle, $\theta_c$, it is convenient in what follows to consider only the $E^{\prime}$ and 
$\theta$ dependence of the acceptance averaged over the vertex coordinates.

Using a model of the spectrometer, the fractional acceptance, 
$A(E^{\prime},\theta)$, is 
calculated by generating Monte Carlo events and taking the ratio of the number of detected events to the number of 
generated events for each bin in phase space.  That is
\begin{equation}
A(E^{\prime},\theta) \equiv N_{acc}(E_{gen}^{\prime},\theta_{gen})/N_{gen}(E_{gen}^{\prime},\theta_{gen}),
\label{eq:accpt}
\end{equation}
where $N_{gen}(E_{gen}^{\prime},\theta_{gen})$ is the number of events generated and 
$N_{acc}(E_{gen}^{\prime},\theta_{gen})$ is the number of events accepted in a given $(E_{gen}^{\prime},\theta_{gen})$ 
bin.  The $gen$ subscripts denote that the kinematics used for the binning are as generated.
The fractional acceptance as defined here is simply a probability.  However, it is evident that $A(E^{\prime},\theta)$ 
depends upon the solid angle, $\Delta \Omega_{gen}(\theta)$, into which events are generated. 
The ``effective'' solid angle coverage for each 2-dimensional bin is 
\begin{equation}
\Delta\Omega_{eff}(E^{\prime},\theta) \equiv A(E^{\prime},\theta)\cdot\Delta\Omega_{gen}(\Delta \theta),
\end{equation}
and is independent of the size of $\Delta\Omega_{gen}(\Delta \theta)$. 
For example, increasing the generation limits of the out-of-plane angle, $X^{\prime}$, from $\pm 100$ mrad to 
$\pm 150$ mrad will decrease  $A(E^{\prime},\theta)$ since $\left | X^{\prime} \right | > 100$~mrad is already 
outside of the collimator aperture.  However, $\Delta\Omega_{gen}(\Delta \theta)$ will increase accordingly and 
$\Delta\Omega_{eff}$ will remain unchanged.

We note that the determination of $A(E^{\prime},\theta)$ does not require generating the events uniformly 
provided that the number generated in each part of phase space is known.  The assumption here is that events 
generated in a given 
$(E^{\prime},\theta)$ bin are not detected in another $(E^{\prime},\theta)$ bin. The fractional acceptance as 
defined is then simply the probability that an event generated in a given bin will be detected in that bin, 
and, therefore, the correction to the yield due to the fractional acceptance in each bin is 
$1/A(E^{\prime},\theta)$.  
If the bin-to-bin migration is small then it is already approximately accounted for 
by redefining the acceptance in Equation~\ref{eq:accpt} to 
\begin{equation}
A(E^{\prime},\theta) = N_{acc}(E_{rec}^{\prime},\theta_{rec})/N_{gen}(E_{gen}^{\prime},\theta_{gen}),
\label{eq:accpt2}
\end{equation}  
where the $rec$ subscripts denote the kinematics as reconstructed.  The $\Delta\Omega_{eff}$ distribution 
extracted from the HMS model for $E^{\prime}$ = 2.8 GeV  and $\theta_c = 12.5^{\circ}$ is shown in 
figure~\ref{accpt}.  The shape in $\theta - \theta_c$ is dominated by the octagonal collimator, which largely 
determines the HMS solid angle acceptance. 
We note that the acceptance is not symmetric in the full scattering angle
when the out-of-plane angle contributes significantly (i.e. at forward in-plane spectrometer angles), even 
though the HMS has a high degree of symmetry about the in-plane scattering angle.  This is because any out-of-plane 
angle will always result in a larger full scattering angle.   

The solid angle defined by the HMS collimator is about 6.75 msr for a point target.  This is slightly reduced for 
a 4 cm extended target and the reduction becomes larger as the spectrometer is moved to larger angles.  At the 
smallest angle measured of $\theta_c = 12.5^{\circ}$, the average solid angle acceptance due to the collimator for 
a momentum bite of $\left | \delta p/p \right | < 8\%$ was determined from the HMS model to be 6.714 msr.  The 
reduction due to all other apertures resulted in a further reduction of only 2.5\% to 6.612 msr.  At the largest 
angle measured of $\theta_c = 80^{\circ}$, the average solid angle acceptance due to the collimator was determined to 
be 6.685 msr, with a further reduction due to other apertures of 5.2\% to 6.335 msr.  For this momentum bite the 
largest reduction of events after the collimator is in the second quadrupole.  

The normalization uncertainty on the acceptance corrections was estimated by combining in quadrature an 
uncertainty of 0.7\% stemming from the reduction in solid angle due to apertures other than the collimator 
(more than one fourth the total at $\theta_c = 12.5^{\circ}$) and an uncertainty of 0.4\% due to the modeling 
of the HMS optics. 

The optical properties of the HMS have been well studied~\cite{dutta} utilizing several techniques 
and a large amount of dedicated optics data taken during many experiments over nearly a decade. 
For the HMS, the optical transport of charged particles through the spectrometer is independent of the momentum 
setting to a very high degree.  The $\Delta \Omega$ distribution at a given $\theta_c$ is then only dependent on 
the energy setting through the dependence of the resolution (including energy straggling) and multiple scattering 
effects in the spectrometer.

\subsection{Elastic Peak Integration}
\label{tail}

\begin{figure}
\includegraphics[width=8.5cm,height=9cm]{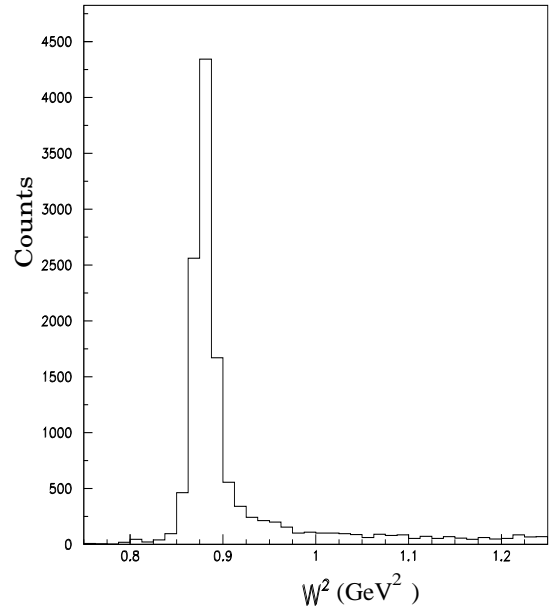}%
\caption{\label{wint_pub}Sample $W^2$ count distribution measured for elastically scattered electrons in a single 
$\theta$ bin after subtraction of the Al quasielastic contribution determined from the dummy target.}
\end{figure}

As already noted, the scattering energy distribution of the elastic peak at an individual $\theta$ 
value is broadened from the $\delta$ function expected in the OPEA due to several effects.  
These include energy resolution effects, and energy loss due to both ionization and bremsstrahlung 
emission.  A typical peak distribution for a single $\theta$ bin is shown in Fig.~\ref{wint_pub}.  
The lower limit of integration is chosen to both minimize the loss of events due to resolution smearing 
and to minimize the sensitivity to potential backgrounds, while the upper limit is chosen to include as 
much of the peak as possible yet to be below the threshold for inelastic $\pi$ production at 
$W^2 \approx 1.16$ $\rm GeV^2$.  

The sensitivities to both the lower and upper limits were studied and were found to be small.  For 
the lower limit, the insensitivity indicates that the aluminum background subtractions are correctly 
handled.  For the upper limit, it indicates that both the resolution effects and the shape of the 
bremsstrahlung distribution are accounted for reasonably well.  Once the upper limit has been chosen, the 
fraction of the distribution that is outside this limit is accounted for by the correction 
factor RC($W^2_{max}$).  If the effects of bremsstrahlung, energy straggling, and resolution are well 
understood, then a corresponding peak integration should be independent of the energy cutoff chosen, 
once the corresponding radiative correction has been applied.   

\begin{figure}
\includegraphics[height=10cm,width=8cm]{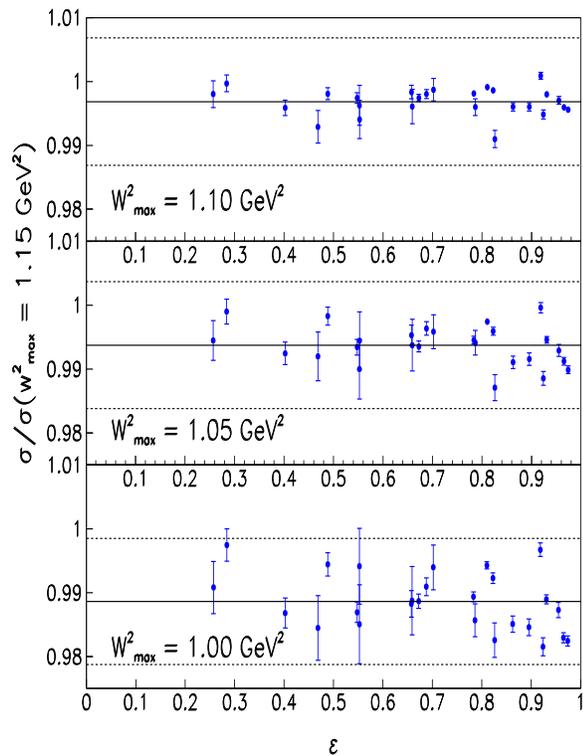}%
\caption{\label{w2cut_pub}(Color online) Ratio of the cross section calculated with a cutoff in the tail 
integration of $W^2$ = 1.10 $\rm GeV^2$ (top), $W^2$ = 1.05 $\rm GeV^2$ (middle), and  $W^2$ = 1.0 $\rm GeV^2$ 
(bottom), relative to a cutoff of $W^2$ = 1.15 $\rm GeV^2$.}
\end{figure}

The sensitivity of the extracted cross section to the energy cutoff is shown in 
Fig.~\ref{w2cut_pub}.  For each kinematic setting, the ratio of the integrated distribution for an 
energy cutoff of $W^2_{max}$ to that for a cutoff of $1.15$ $\rm GeV^2$ is plotted versus $\varepsilon$.  
In the upper plot $W^2_{max} = 1.10$ $\rm GeV^2$, in the middle plot $W^2_{max} = 1.05$ $\rm GeV^2$, and 
for the bottom plot $W^2_{max} = 1.0$ $\rm GeV^2$.  
The typical point-to-point difference in tail-corrected integration is less than $0.3\%$ for the largest 
change in the cutoff value and shows little $\varepsilon$ dependence.  We take this as the estimated random 
point-to-point uncertainty on this procedure and include it in the uncertainty of the radiative corrections.  
Additionally, we note that the normalization difference of about 1\% between between the smallest and 
largest values for $W^2_{max}$ is likely due to a combination of unoptimized resolution matching and the 
approximate handling of the energy straggling in the simulation used for generating the acceptance corrections.  
However, this optimization becomes much less important as more of the peak is integrated.  We take 0.35\% as the 
estimated normalization uncertainty on this procedure and include it in the uncertainty of the radiative corrections.    

Finally, we note that, except for the three measurements at beam energies below 2 GeV, the same 
$W^2_{max}$ value was used for all kinematic settings.  This was possible because of the large 
acceptance of the HMS spectrometer.  This is in contrast to previous precision 
measurements~\cite{walker,andiv}, in which the spectrometer acceptance determined the maximum 
$W^2_{max}$ at each kinematic setting.
 

\subsection{$\theta$ Bin-Centering and Averaging}

After performing the peak integration, the cross section is then extracted for each $\theta$ bin.  
Often, the statistics taken in each bin are small ($<$ 2000 counts).  In order to improve the 
statistical accuracy, one would like to combine the data from all $\theta$ bins.  If the cross 
section did not depend (or depended only linearly) on the scattering angle, the cross sections 
extracted in each bin could simply be averaged.  This is not the case, however.  The HMS spectrometer 
has a relatively large acceptance of about $\pm$1.8 degrees in the scattering angle.  Therefore, the cross 
section can vary greatly across the angular acceptance.  At some kinematics, this variation can be a 
factor of 3 or more (and strongly non-linear) across the acceptance.  In order to average the cross 
sections in each $\theta$ bin, the $\theta$ dependence of the cross section must be corrected for.  

This correction is called ``$\theta$ bin-centering'' (BC), and our prescription for it is 
straightforward.  Since we would like to quote the cross section at the central angle of the spectrometer, 
the following correction is applied to each $\theta$ bin:
\begin{equation}
\left[\frac{d\sigma(\theta)}{d\Omega}\right]_{BC,i} = \frac{d\sigma(\theta_i)}{d\Omega}
\cdot \frac{\sigma^{Mod}(\theta)}{\sigma^{Mod}(\theta_i)},
\end{equation}
where $\theta$ is the central angle, $\theta_i$ is the angle for the $i^{th}$ bin, and 
$\sigma^{Mod}$ is the value of a cross section model.  For this procedure to be valid, care must 
be taken to subtract all backgrounds and to apply all corrections that have a $\theta$ dependence, 
bin-by-bin.  This includes radiative corrections.  The bin-centered cross sections can then be 
averaged over the $\theta_i$ to give the measured cross section at the central spectrometer angle.  
This was done as a weighted average, where the inverse of the square of the full statistical errors 
was used as a weighting factor.  The statistical errors take into account the statistics of 
both the hydrogen and subtracted target endcap events, as well as the acceptance correction uncertainties 
due to statistical errors on the Monte Carlo generation.        

\begin{figure}
\includegraphics[height=10cm,width=8cm]{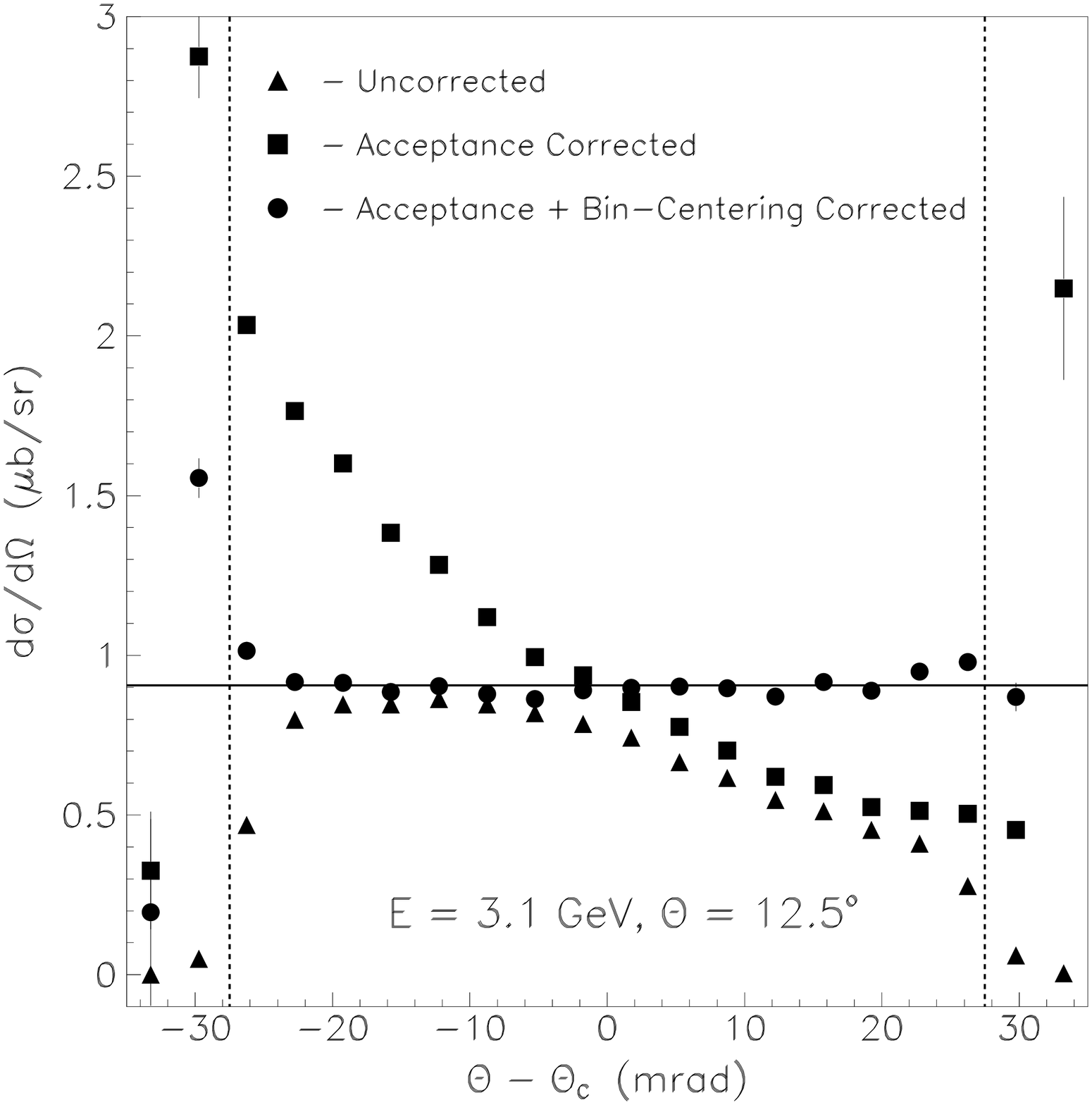}%
\caption{\label{method_pub}Cross section extracted in each $\theta$ bin across the HMS acceptance.  
The various data are discussed in the text.}
\end{figure}

An example of this procedure is presented in Fig.~\ref{method_pub}.  Shown is the cross section 
extracted at a beam energy of 3.12 GeV and a central HMS angle of $12.5^{\circ}$, before both acceptance 
and BC corrections (triangles).  Also plotted is the cross section after applying acceptance 
corrections (squares) and after applying both acceptance and bin-centering corrections (circles).  
Only statistical uncertainties in the data are included in the error bars shown.  However, the 
calculated acceptance corrections for bins at the edge of the acceptance can have large fractional 
errors, as they are very sensitive to both accurate modeling of the multiple scattering processes and 
small variations in the positions of apertures like the collimator.  To minimize the effects of such 
sensitivities, bins at the edge of the $\theta$ acceptance where the calculated acceptance was below 
some minimum value were neglected in the averaging procedure.  The angular acceptance limits used in the 
present experiment are represented by the vertical dashed lines in Fig.~\ref{method_pub}.  The cross 
sections obtained after averaging over these limits were found to be quite insensitive to the effects 
described above.  Uncertainties associated with these effects were studied by adjusting aperture/target 
positions, magnet fields, and multiple scattering distributions within reasonable limits and determining 
the corresponding acceptance function.  The cross section extracted with this acceptance function was 
typically found to agree with that using the nominal acceptance function to within 0.5\%.    

The shape of the uncorrected distribution is the convolution of the $\theta$ dependence of the cross 
section with the acceptance of the spectrometer, with the latter being primarily defined by the 
collimator.  Correcting for the acceptance leaves only the $\theta$ dependence, which is then removed 
by the bin-centering corrections.  The resulting, fully corrected, cross section should then be a 
constant across the $\theta$ acceptance and equal to the cross section at the central $\theta$ to 
within statistical fluctuations.  This is indeed the case, allowing for small variations which are 
mostly due to an imperfect optics model for the spectrometer, as well as the approximate treatment 
of the multiple scattering effects in the simulations.

\subsection{Radiative Corrections}
\label{sect:rc}

\begin{figure}
\includegraphics[height=8cm,width=8cm]{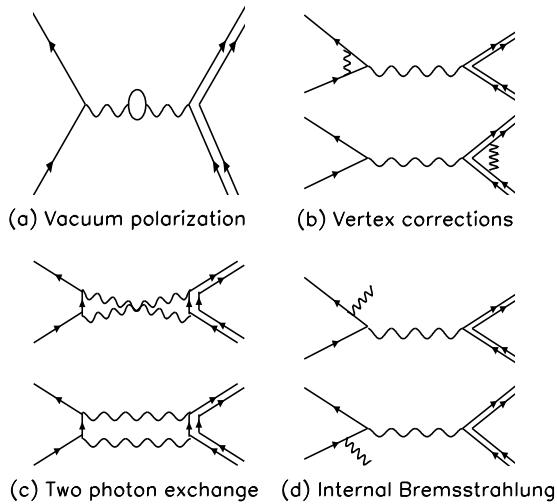}
\caption{\label{raddiag} Feynman diagrams for higher order QED processes, including (a) vacuum 
polarization, (b) vertex corrections, (c) two photon exchange, and (d) internal bremsstrahlung 
emission from electrons.  Not shown are diagrams for proton bremsstrahlung, or external 
bremsstrahlung.}
\end{figure}
While the form factors can be easily extracted from the Born cross section for single photon exchange, 
the cross section that is measured in a scattering experiment includes higher order electromagnetic 
processes which are depicted in Fig.~\ref{raddiag}.  These processes can be categorized into two 
types:  1) the internal processes, which originate due to the fields of the particles at the scattering 
vertex, and 2) the external processes, which originate due to the fields of particles in the bulk target 
materials.  The internal processes include: vacuum polarization, vertex corrections, two-photon exchange, 
and (internal) bremsstrahlung emission in the field of the proton from which the scattering took place, 
while the external process is due to (external) bremsstrahlung in the field of a proton in the material 
either before or after the scattering vertex.  

The radiative correction factors, which account for these higher order processes, were calculated using 
the same procedure as the high precision SLAC data~\cite{walker,andiv} and discussed in detail in 
Reference~\cite{walker}.  This procedure is based on the prescription of Mo and Tsai~\cite{tsai,motsai}.  
The radiative correction (RC) is applied to the measured cross section (after integration 
of the bremsstrahlung tail) as a multiplicative factor, i.e., 

\begin{equation}
\left [\frac{d\sigma(\theta)}{d\Omega}\right ]_{Born} = RC \cdot \left [\frac{d\sigma(\theta)}{d\Omega}\right ]_{Meas}.
\end{equation}  

The corrections calculated include only the infrared divergent contributions from the two-photon exchange 
and proton vertex diagrams, with the non-divergent contributions having been previously estimated~\cite{motsai} 
to be less than 1$\%$.  However, the topic of two-photon exchange has recently been of renewed theoretical 
interest~\cite{2pho1,2pho2,2pho3} in light of the discrepancy between the elastic form factor ratios extracted 
from Rosenbluth separations of cross section measurements and those measured in polarization transfer experiments.  
In addition, a recent study~\cite{positrons} of the world's data on the ratio of elastic cross sections for 
$e^+p$ to $e^-p$ has recently been made to look for evidence of two-photon effects. 

The radiative corrections applied at each kinematic setting are listed in Table~\ref{cstable}.  We note that 
these correction factors are significantly smaller than those applied in References~\cite{walker,andiv}.  
This is mostly due to integrating more of the radiative tail, but also to the reduction of external bremsstrahlung 
afforded by our much shorter target.  

The uncertainties in the radiative correction procedure were studied in 
References~\cite{walker,andiv} and were estimated to be 0.5\% point-to-point and 1.0\% nomalized.  To 
these we have added, in quadrature, the additional contributions resulting from the radiative tail integration, 
which are discussed in Section~\ref{tail}. 
. 

\subsection{Additional Corrections Applied to Large $E^{\prime}$ Data}

The four cross section measurements at scattering energies greater than 3.5 GeV (indicated with a $\star$ in 
Table~\ref{cstable}) required two additional corrections that were not needed for the rest of the data set.  
These corrections, which will be discussed in what follows,  account for the following two effects:  1) an 
additional change in the effective target density, and 2) a small misfocusing in the spectrometer optics, 
relative to nominal. 
    
During the data taking it was discovered that the fan controlling the flow of hydrogen through the cryogenic 
target had been inadvertently lowered from the 60~Hz nominal speed to 45~Hz.  This resulted in larger localized 
boiling due to the beam and effectively lowered the density of hydrogen.  To account for this, high statistics 
runs ($\approx 0.1\%$) were performed at both fan speeds and the size of the effect was determined from the ratio 
of cross sections.  Since the effect of target boiling was already measured for a fan speed of 60~Hz, an 
additional correction to the yields of $+0.6\% \pm 0.2\%$ was included for the data taken at the lower speed.

In addition to the correction for the difference in target density, the cross section measurements at large $E'$ 
were also corrected for a slight misfocusing of the spectrometer.  A TOSCA~\cite{tosca} model of the HMS dipole 
indicated that it would start to suffer from saturation effects starting at currents corresponding to a momentum 
of 3.5 GeV for the central ray and that this effect would increase quadratically with momentum.  This correction 
was included when setting the current in the dipole for this experiment, but was later found from both kinematic 
and optics studies to not be needed.  These studies indicated that the actual HMS field continues to increase 
linearly with current to a high accuracy up to the highest momentum tested of 5.1 GeV.   

The missetting of the dipole field due to this unneeded saturation correction resulted in reconstructing the 
wrong scattering angle by an amount that varied across the angular acceptance, and led to a depletion of events 
that reconstructed in the region of the angular acceptance used in the analysis.  The correction to the 
reconstructed scattering angle was determined by requiring $W = M_p$ for each $\theta$ bin, and this correction 
was then fit as a function of $\theta$ across the angular acceptance.  The result of this fit was applied as a 
correction to the scattering angle event-by-event, and effectively reshuffled events back into the depleted bins.      
The effect of this correction on the cross section was largest at the highest scattering momentum in this experiment 
of 4.7 GeV and resulted in a $1.6\%$ increase.  The uncertainty on this correction was estimated to be $0.4\%$ and 
was assumed to be the same at all four kinematics where a correction was applied.

\section{Systematic Uncertainties}

The estimated systematic uncertainties for the experiment are listed in Table~\ref{syst} for those that are 
assumed random point-to-point in $\varepsilon$ and in Table~\ref{syst_norm} for those that effect the overall 
normalization uncertainty only.  The quadrature sum of the point-to-point and normalization uncertainties gives 
the absolute uncertainties on the cross section measurements.  The point-to-point uncertainties are those that 
depend upon variable run conditions or kinematics.   Discussions of the uncertainties presented here can be found 
in earlier sections of the text.

\begin{table}[tbh]
\caption{E94-110 point-to-point systematic uncertainties.}
\begin{center}
\begin{tabular}{l l l}
\hline
\hline
 Experimental Quantity & Uncertainty      &   $\Delta \sigma/\sigma$  \\
                       &                  &        (pt-pt)            \\
\hline
Beam Energy            & $4\cdot10^{-4} $ & 0.0024                    \\
Scattering Angle       &  0.2 mrad        & 0.0026                    \\ 
Target Density         &  0.1\%          & 0.001                    \\ 
Target Length          &  0.1\%           & 0.001                     \\
Beam Charge            & 0.2\%            & 0.002                     \\ 
Acceptance             & 0.5\%            & 0.005                     \\
Detector Efficiency    & 0.15\%           & 0.0015                     \\
Tracking Efficiency    & 0.25\%           & 0.0025                    \\
Deadtime Corrections   & 0.14\%           &  0.0014                   \\
Target Cell Background & 0.2\%            &  0.002                    \\  
Radiative Corrections  & 0.6\%            &  0.006                    \\
\hline
                       & Total            & 0.0097                    \\
\hline
\hline
\end{tabular}
\label{syst}
\end{center}
\end{table}
\begin{table}[tbh]
\caption{E94-110 normalization uncertainties.}
\begin{center}
\begin{tabular}{l l l}
\hline
\hline
 Experimental Quantity & Uncertainty        &   $\Delta \sigma/\sigma$    \\
                       &                    &        (Norm)               \\
\hline
Beam Energy            & $4\cdot10^{-4} $   & 0.0024                      \\
Scattering Angle       &  0.4 mrad          & 0.0053                      \\
Target Density         &  0.4\%             & 0.004                       \\
Target Length          &  0.3\%             & 0.003                       \\ 
Beam Charge            &  0.33\%            & 0.0033                      \\ 
Acceptance             & 0.8\%              & 0.008                       \\
Detector Efficiency    & 0.4\%              & 0.004                       \\
Tracking Efficiency    & 0.3\%              & 0.003                       \\
Deadtime Corrections   & 0.1\%              &  0.001                      \\
Target Cell Background & 0.3\%              &  0.003                      \\
Radiative Corrections  & 1.1\%              &  0.011                      \\
\hline
                       & Total              & 0.017                       \\
\hline
\hline
\end{tabular}
\label{syst_norm}
\end{center}
\end{table}

\section{Results}

\begin{center}
\begin{table*}[tbch]
\caption{Table of measured {\it ep} elastic cross sections.  The systematic uncertainties listed are the 
estimated point-to-point uncertainties for each kinematic setting.  In addition there is scale uncertainty 
of 1.7\%.}
\begin{tabular}{ c  c  c  c  c  c  c  c }
\hline
\hline
 \hspace{1cm}  &  \hspace{1cm}  &  \hspace{1.2cm}  &  \hspace{1.2cm}  &  \hspace{1.8cm}   &  \hspace{1.8cm}  & \hspace{1.8cm}  &  \hspace{1cm}     \\
 $E_{beam} $  &  $\Theta$  & $Q^2$ & $\varepsilon$  & $\sigma$ & $\Delta \sigma$ (stat) & 
 $\Delta \sigma$ (sys) &  RC \\
    (GeV)        &  (Deg)  &  $({\rm GeV}/c)^2$      &   &    ($\mu$b/sr)  &   ($\mu$b/sr)   &   ($\mu$b/sr)      &      \\
\hline
1.148 & 47.97 & 0.6200 & 0.6824 & 0.1734E-01 &   0.35E-04 &   0.16E-03 & 1.055  \\
1.148 & 59.99 & 0.8172 & 0.5492 & 0.4813E-02 &   0.11E-04 &   0.45E-04 & 1.047  \\
1.882 & 33.95 & 0.8995 & 0.8104 & 0.1464E-01 &   0.28E-04 &   0.14E-03 & 1.098  \\
2.235 & 21.97 & 0.6182 & 0.9187 & 0.1098E+00 &   0.40E-03 &   0.11E-02 & 1.120  \\
2.235 & 31.95 & 1.1117 & 0.8226 & 0.8938E-02 &   0.25E-04 &   0.86E-04 & 1.121  \\
2.235 & 42.97 & 1.6348 & 0.6879 & 0.1184E-02 &   0.54E-05 &   0.11E-04 & 1.114  \\
2.235 & 58.97 & 2.2466 & 0.4885 & 0.1522E-03 &   0.86E-06 &   0.14E-05 & 1.100  \\
2.235 & 79.97 & 2.7802 & 0.2843 & 0.2868E-04 &   0.21E-06 &   0.27E-06 & 1.083  \\
3.114 & 12.47 & 0.4241 & 0.9740 & 0.8968E+00 &   0.25E-02 &   0.96E-02 & 1.150  \\
3.114 & 15.97 & 0.6633 & 0.9553 & 0.1892E+00 &   0.80E-03 &   0.20E-02 & 1.151  \\
3.114 & 19.46 & 0.9312 & 0.9308 & 0.4802E-01 &   0.12E-03 &   0.49E-03 & 1.154  \\
3.114 & 32.97 & 2.0354 & 0.7835 & 0.1017E-02 &   0.26E-05 &   0.99E-05 & 1.154  \\
3.114 & 40.97 & 2.6205 & 0.6726 & 0.2125E-03 &   0.74E-06 &   0.20E-05 & 1.150  \\
3.114 & 49.97 & 3.1685 & 0.5480 & 0.5568E-04 &   0.28E-06 &   0.53E-06 & 1.147  \\
3.114 & 61.97 & 3.7261 & 0.4026 & 0.1487E-04 &   0.10E-06 &   0.14E-06 & 1.137  \\
3.114 & 77.97 & 4.2330 & 0.2574 & 0.4260E-05 &   0.49E-07 &   0.40E-07 & 1.121  \\
4.104 & 38.97 & 3.7981 & 0.6578 & 0.4919E-04 &   0.31E-06 &   0.47E-06 & 1.185  \\
4.104 & 45.96 & 4.4004 & 0.5528 & 0.1582E-04 &   0.29E-06 &   0.15E-06 & 1.181  \\
4.413 & 44.98 & 4.7957 & 0.5526 & 0.1098E-04 &   0.20E-06 &   0.11E-06 & 1.192  \\
4.413 & 50.99 & 5.2612 & 0.4686 & 0.4898E-05 &   0.73E-07 &   0.47E-07 & 1.188  \\
$\star$5.494 & 12.99 & 1.3428 & 0.9655 & 0.4055E-01 &   0.11E-03 &   0.50E-03 & 1.214  \\
$\star$5.494 & 17.96 & 2.2878 & 0.9239 & 0.2866E-02 &   0.13E-04 &   0.33E-04 & 1.220  \\
$\star$5.494 & 20.47 & 2.7822 & 0.8955 & 0.9824E-03 &   0.41E-05 &   0.11E-04 & 1.223  \\
$\star$5.494 & 22.97 & 3.2682 & 0.8627 & 0.3770E-03 &   0.15E-05 &   0.42E-05 & 1.226  \\
5.494 & 25.47 & 3.7385 & 0.8261 & 0.1608E-03 &   0.14E-05 &   0.18E-05 & 1.226  \\
5.494 & 27.97 & 4.1867 & 0.7865 & 0.7749E-04 &   0.61E-06 &   0.77E-06 & 1.228  \\
5.494 & 32.97 & 5.0031 & 0.7023 & 0.2149E-04 &   0.23E-06 &   0.21E-06 & 1.226  \\
5.494 & 35.48 & 5.3699 & 0.6593 & 0.1267E-04 &   0.20E-06 &   0.12E-06 & 1.225  \\
\hline
\hline
\end{tabular}
\label{cstable}
\end{table*}
\end{center}

The complete set of Born cross sections extracted from these precision elastic scattering measurements 
are listed in Appendix A.  In total, measurements at 28 different kinematics were included in 
the final data set, covering a range in $Q^2$ from approximately 0.4 GeV to 5.5 GeV.  In addition, 
a significant range in $\varepsilon$ was covered, even though the elastic kinematics were 
not specifically optimized for Rosenbluth separations.

\subsection{Comparisons to Fits of the Existing World's Data}

\begin{figure}
\includegraphics[height=14cm,width=8cm]{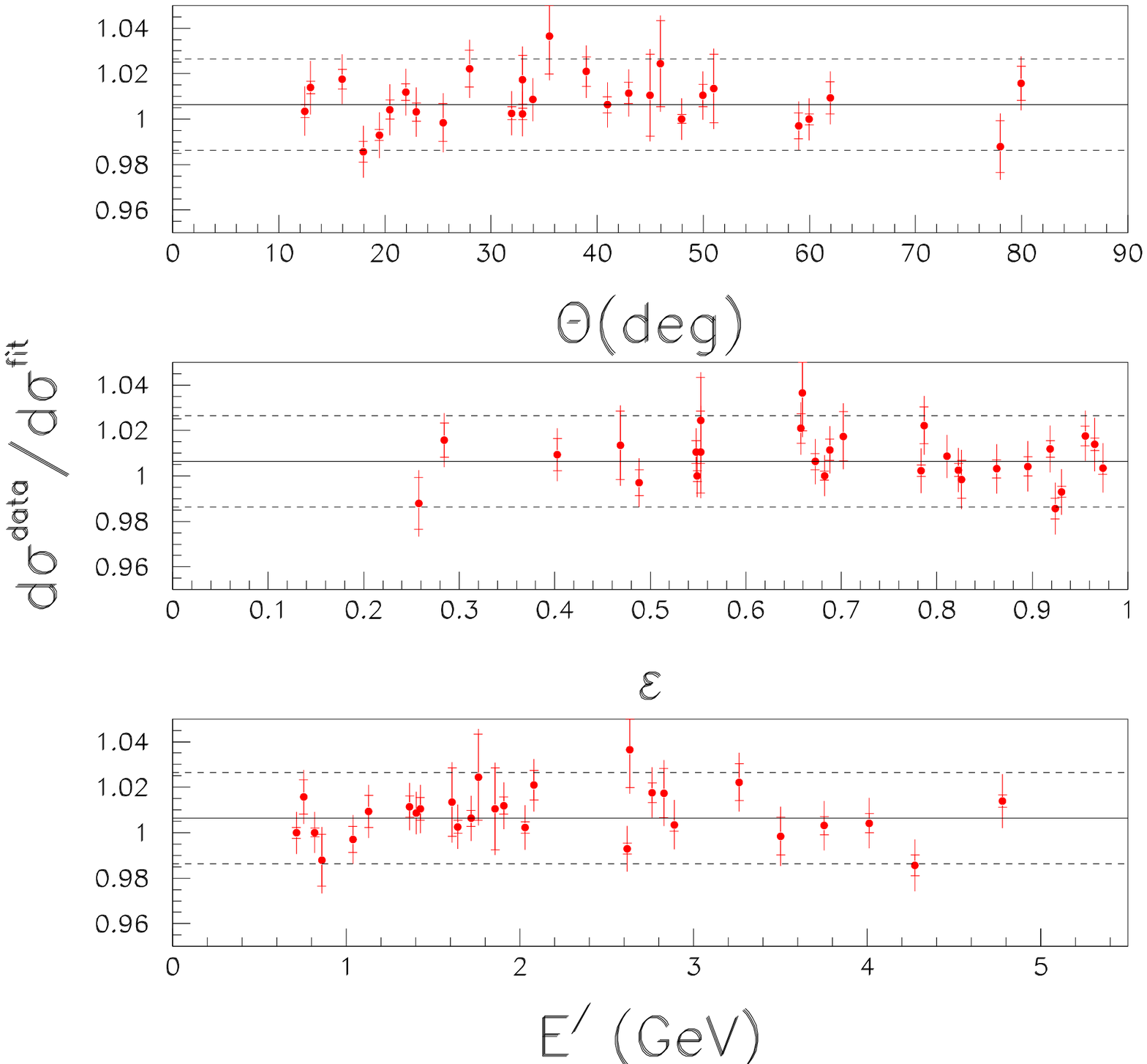}
\caption{(Color online) The ratio of measured {\it ep} elastic cross sections to the Arrington fit of 
previous cross section data, as a function of scattering angle, longitudinal photon polarization, and 
scattering energy.  The solid line indicates the average ratio of 1.006 and the dashed lines indicate 
$\pm 2 \%$ of this value.}
\label{ratio3}
\end{figure}

Comparisons of these cross sections were made to recent fits of the previous world's data set.  
These included the fit of Brash $et$ al. from~\cite{brash} and the fits of Arrington from~\cite{arrington}.  
Both analyzes performed a fit to the combined cross section and polarization transfer data, 
while that of Arrington also included a fit to the cross section data alone.  It should be stressed 
that, although nearly the same data sets were included in the fits, the method of combining the 
polarization transfer data differed.

In the work of Brash $et$ al., the polarization transfer results for the $Q^2$ dependence of $G_E/G_M$ were used 
as a constraint to do a one-parameter refit of the Rosenbluth data, from which $G_M$ was extracted.  
A $\chi^2$ minimization was then performed to fit $G_M$ as a function of $Q^2$.  In the work of Arrington, 
two fits were performed: one that included the polarization transfer data, and one that did not.  
However, here the polarization transfer results for $G_E/G_M$ were included with the cross section 
measurements as equally weighted points in a $\chi^2$ minimization to fit both form factors 
simultaneously.  This follows closely the older work of Walker~\cite{walker} on global fits to cross 
section data.

The ratio of the data to the Arrington fit of cross section data is shown in Fig.~\ref{ratio3} versus 
$\theta$, $\varepsilon$, and $E'$.  The inner error bars represent the purely statistical uncertainties, 
while the full error bars include the point-to-point uncertainties as well.  This fit is observed to 
describe the data set very well over the entire kinematic range.  The $\chi^2$ (calculated using 
only the point-to-point uncertainties and after 
removing the average $0.6\%$ normalization difference between the current data set and the previous 
world data set of cross section measurements) 
distribution was found to be well described by a Gaussian distribution with a width corresponding to an 
average uncertainty of about 1.0$\%$, consistent with the estimated errors combining the systematic 
point-to-point and statistical uncertainties in quadrature.  For each of the three fits previously 
described, the total $\chi^2$ per degree of freedom ($\chi^2_{\nu}$) to the data was calculated.  The 
results for the region above $Q^2$ = 1 $({\rm GeV}/c)^2 $, where the discrepancy between the cross 
section and polarization results differ significantly, were found to be $\chi^2_{\nu} = $ 0.76 
(Arrington fit to cross sections), 1.06 (Arrington fit including polarization transfer results), and 
2.95 (Brash $et$ al fit), allowing the overall normalization to vary.  

These results are interesting for two reasons.  Firstly, the full data set above 
$Q^2$ = 1 $({\rm GeV}/c)^2$ favor the fit to cross section data only over the fits that includes 
the polarization transfer data.  Secondly, the data favor the Arrington prescription for combining 
the cross section and polarization transfer data over the prescription of Brash  $et$ al.  This does not 
resolve the inconsistency between the Rosenbluth and polarization transfer results, but rather 
underscores a consistency in the global cross section data set including these new measurements.  
The discrepancy with these and the polarization transfer measurements is highlighted further.

\subsection{Rosenbluth Extractions of Form Factors}

\begin{figure}
\includegraphics[height=14cm,width=8cm]{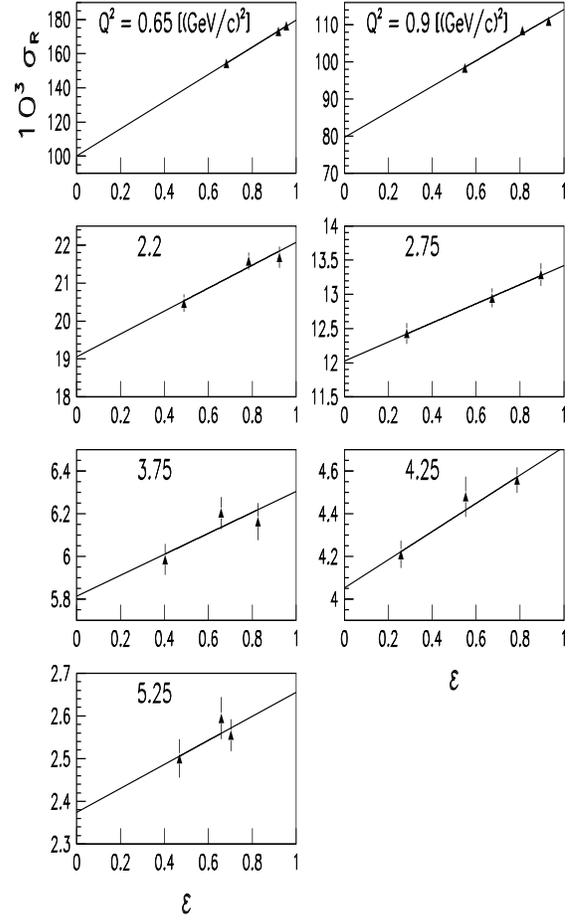}
\caption{Rosenbluth separations of the form factors.  Plotted is the reduced cross 
sections ($\times 1000$) 
versus $\varepsilon$.}
\label{lts}
\end{figure}
The individual Sachs form factors were extracted from the cross section data at seven different $Q^2$ values 
via the Rosenbluth separation method.  This required that the cross section measurements at similar $Q^2$ values 
be grouped.  Since none of the measurements were taken at precisely the same $Q^2$, a correction factor was applied 
to some of the cross sections in each group to evolve to a common $Q^2$.  The correction factor for this $Q^2$ 
evolution was calculated from fits to previous data via
\begin{equation}
\frac{d\sigma(Q^2_c,\varepsilon)}{d\Omega} = \frac{d\sigma(Q^2,\varepsilon)}{d\Omega}
\cdot \frac{\sigma^{Mod}(Q^2_c,\varepsilon)}{\sigma^{Mod}(Q^2,\varepsilon)},
\end{equation}
where $\sigma^{Mod}$ is the value given by the fit of Arrington to the cross sections data, and $Q^2$, $Q^2_c$, 
represent the values before and after the evolution, respectively. 

In order to perform Rosenbluth separations at a particular $Q^2$, the following two conditions on the 
data were required:  1) each separation must contain three distinct $\varepsilon$ points, and 2) the $Q^2$ 
evolution for each $\varepsilon$ point must constitute less than a 15$\%$ correction.   
The sensitivity of the extracted form factors on the model used for the $Q^2$ evolution was found to be much 
less than the uncertainties.  Plots of the reduced cross sections 
versus $\varepsilon$ are presented in Fig.~\ref{lts} for each $Q^2$.  Also presented are the results of the 
linear fit.  The error bars on each point represent the total point-to-point uncertainties, including both 
statistical and systematic uncertainties added in quadrature.

\
\begin{center}
\begin{table*}[tbch]
\caption{Table of the Rosenbluth extracted Sachs form factors relative to the dipole form factor, 
$G_d = \frac{1}{(1 + Q^2/0.71)^2}$ ($Q^2$ in $({\rm GeV}/c)^2$.}
\begin{tabular}{ c  c  c  c}
\hline
\hline
\hspace{1.2cm}       &  \hspace{2.4cm}   &  \hspace{2.4cm}   & \hspace{2.4cm}            \\
  $Q^2$              & ${G_M}_p /(\mu_p  G_{dip})$         &  ${G_E}_p / G_{dip}$        &  $\mu {G_E}_p / {G_M}_p$  \\
  $({\rm GeV}/c)^2$  &                   &                   &                           \\
\hline
 0.65 & 0.968 $\pm$ 0.032 & 1.035 $\pm$ 0.052 & 1.069 $\pm$ 0.085  \\
 0.91 & 1.028 $\pm$ 0.019 & 0.954 $\pm$ 0.053 & 0.928 $\pm$ 0.067  \\
 2.20 & 1.050 $\pm$ 0.016 & 0.923 $\pm$ 0.121 & 0.878 $\pm$ 0.125  \\
 2.75 & 1.055 $\pm$ 0.010 & 0.888 $\pm$ 0.114 & 0.841 $\pm$ 0.109  \\
 3.75 & 1.044 $\pm$ 0.015 & 0.873 $\pm$ 0.232 & 0.837 $\pm$ 0.220  \\
 4.20 & 1.012 $\pm$ 0.012 & 1.255 $\pm$ 0.157 & 1.240 $\pm$ 0.163  \\
 5.20 & 1.007 $\pm$ 0.032 & 1.183 $\pm$ 0.511 & 1.176 $\pm$ 0.552  \\
\hline
\hline
\end{tabular}
\label{fftable}
\end{table*}
\end{center}

\begin{figure}
\includegraphics[height=11cm,width=8.5cm]{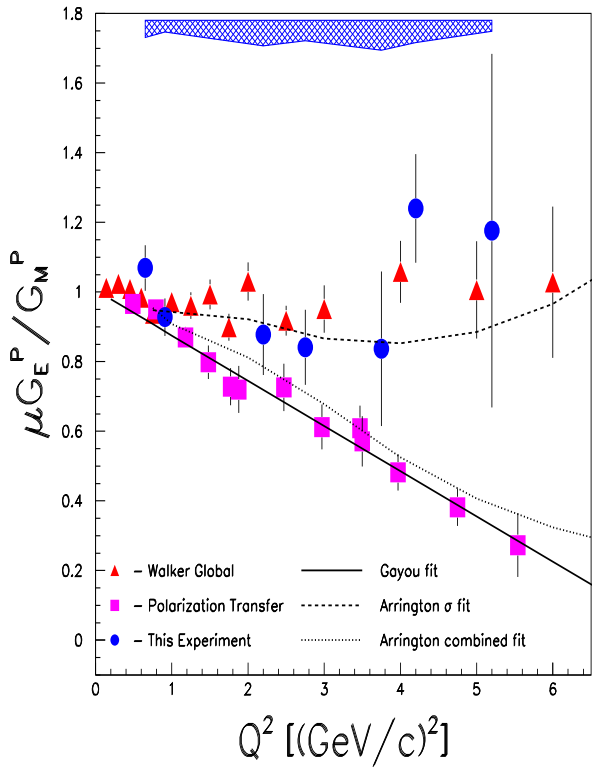}
\caption{(Color online) Extracted values of $\mu {G_E} / {G_M}$.  Also shown are those extracted from previous 
Rosenbluth and polarization transfer measurements.  The error bars on the data include the full point-to-point 
uncertainties, while the hatched band at the top indicates the uncertainty due to the absolute uncertainty 
in the scattering angle of 0.4 mrad.}
\label{gegm_new}
\end{figure}

The results for the ratio $\mu {G_E} / {G_M}$ extracted from the current data set are presented in 
Fig.~\ref{gegm_new}, along with previous extractions from both cross section~\cite{walker} and 
polarization transfer~\cite{jones,gayou1} data.  The current data are seen to agree well with the previous 
cross section data, while being in significant disagreement with the polarization transfer results.  The error 
bars on each point represent the uncertainties obtained for the fit parameters, while the hatched band at the 
top of the figure represents that due to the estimated 0.4 mrad uncertainty in the absolute scattering angle.  
To a large degree, an error in the scattering angle would shift the entire set of ratios up or down but would 
not significantly alter the trend versus $Q^2$, which is significantly different from that of the polarization 
transfer data.  

\section{Conclusion}

We have performed high precision measurements of the {\it ep} elastic cross section covering a considerable amount of 
the $Q^2 - \varepsilon$ space for which there exists a large discrepancy between Rosenbluth and polarization transfer 
extractions of the ratio $\mu {G_E} / {G_M}$.  This data set shows good agreement with previous cross section 
measurements, indicating that if a here-to-fore unknown systematic error does exist in the cross section measurements 
then it is intrinsic to all such measurements.  

A likely candidate, which has received much theoretical interest 
recently~\cite{2pho1,2pho2,2pho3}, is possible contributions from two-photon exchange, which are not fully accounted for 
in the standard radiative corrections procedure of Mo-Tsai.  Although it is currently unclear whether such an effect 
can fully explain the discrepancy, considerable progress is being made.  

Complementary to this theoretical effort is the recently completed experiment~\cite{super} in JLab Hall A which utilizes 
the so-called `Super-Rosenbluth' technique to extract the form factor ratio.  This experiment measured the proton cross 
sections and is therefore sensitive to a different set of systematic uncertainties than the previous electron cross section 
data.  However, these measurements are still as sensitive to two-photon exchange effects as electron cross section 
measurements and will, therefore, provide a vital clue whether such effects are present.     
In any event, it is critical that the source of the discrepancy be found if there is to be any hope of extracting the $Q^2$ 
dependence of the individual form factors.

\begin{acknowledgments}
This work was supported in part by research grants 0099540 and 9633750 from the National Science Foundation and 
under contract W-31-109-ENG-38 from the U.S. Department of Energy, Nuclear Physics Division.  We are 
grateful for the outstanding support provided by the Jefferson Lab Hall C scientific and engineering staff.  The 
Southeastern Universities Research Association operates the Thomas Jefferson National Accelerator Facility under the 
U.S. Department of Energy contract DEAC05-84ER40150.
\end{acknowledgments}


\clearpage


\end{document}